\documentclass[a4paper, 12pt]{article}
\usepackage{amsmath}
\usepackage{amssymb}
\usepackage{psfrag}
\usepackage{latexsym}
\usepackage{enumerate}
\usepackage{graphics}
\usepackage{subfigure}
\usepackage{caption}
\input epsf
\usepackage[dvips]{graphicx}
\usepackage{enumerate}
\usepackage[english]{babel}
\usepackage{amsmath,amssymb,amsthm}

\newcommand{\be}{\begin{equation}}
\newcommand{\ee}{\end{equation}}
\newcommand{\bea}{\begin{eqnarray}}
\newcommand{\eea}{\end{eqnarray}}

\newcommand{\nod}{\noindent}

\newcommand{\ba}{\begin{array}}
\newcommand{\ea}{\end{array}}
\newcommand{\bc}{\begin{center}}
\newcommand{\ec}{\end{center}}

\allowdisplaybreaks

\begin{document}
\title{Dynamical graphs for the SI epidemiological model}
\date{October 2011}
\author{
\small{Jose L. Herrera $^{a}$}\\
\small{Gilberto Gonz\'alez-Parra$^{a}$}\footnote{Corresponding author}\\
\small{ Email:gcarlos@ula.ve}\\
\small{$^{\rm a}${Grupo Matem\'atica Multidisciplinar, Facultad de Ingenier\'ia.}}\\
\small{Universidad de los Andes, M\'erida, Venezuela.}}


\maketitle \thispagestyle{empty}

\begin{abstract}
In this paper we study the susceptible-infectious (SI) epidemiological model using dynamical graphs. Dynamical structures have been recently applied in many areas including complex systems. Dynamical structures include the mutual interaction between the structure topology and the characteristics of its members. Dynamical graphs applied to epidemics consider generally that the nodes are individuals and the links represent different classes of relationships between individuals with the potential to transmit the disease. The main aim in this article is to study the evolution of the SI epidemiological model and the creation of subgraphs due to the dynamic behavior of the individuals trying to avoid the contagious of the disease. The proposed dynamical graph model uses a single parameter which reflects the probability of rewire that represent actions to avoid the disease. This parameter includes also information regarding the infectivity of the disease. The numerical simulations using Monte Carlo method show that the dynamical behavior of individuals affects the evolution of the subgraphs. Furthermore, it is shown that the connectivity degree of the graphs can change the arise of subgraphs and the asymptotic state of the infectious diseases.
\end{abstract}
MSC classes: 62P10 60J10\\
\nod {\bf Keywords:Graph dynamics, SI epidemic model, Coevolution, Adaptative social behavior, Extinction time.}
\newpage

\section{Introduction}

One of the most important issues related to real world is health. Thus, epidemic attracts a lot of attention. The dynamical behavior of epidemic diseases have been studied for a long time, $SI, SIS, SIR$ and $SIRS$ are important and fundamental epidemic models \cite{Hethcote,Murray}. There are several approaches to study epidemics on dynamical structures. One important branch are the models that include interactions between the network topology and the states of the nodes \cite{ThiloAdaptative,ThiloInterface,MarceauAdaptative}.

Other approach is the deterministic, which use generally models that rely on systems of differential equations that represent the number of individuals in various categories through continuous variables. Systems of ordinary differential equations ODE are well-known tools that have been used to model different type of diseases \cite{Hethcote,Murray}. However, the classical continuous models have some drawbacks such that it neglects local characteristics that may affect the evolutionary process. Continuous models have a well developed theoretical foundation and its analysis are easier than in the discrete individual models. Traditional classical mathematical epidemic models assume that the population is fully mixed. In these models describing different populations as continuous variables is a poor approximation in small populations. An alternative model is a Markov process with a discrete population and time, in which the probability of transition from (s, i) to (s -1, i +1) is given by $\beta s i \Delta t$, where $s$ is representing the susceptible individuals and $i$ the infectious ones. This process can be simulated stochastically on computer by using Monte Carlo simulations.

It has been suggested recently that many social, biological, and communication systems possess two universal characters, the small-world effect and the scale-free property \cite{EpidemicLatticeScaleFree}. In view of the wide occurrence of complex networks in nature, it is important to study the effects of topological structures on the dynamics of epidemic spreading \cite{EpidemicLatticeScaleFree,ThiloAdaptative}. An important previous result is that infections with low spreading rates will prevail over the entire population in scale-free networks, which radically changes many of the conclusions drawn in classic epidemic modelling \cite{EpidemicLatticeScaleFree,EpidemicScaleFree}.

The classical SI epidemiological model has been proposed and applied to the circulation of the Feline Immunodeficiency Virus (FIV) within domestic cats. FIV is a lentivirus which is structurally similar to the Human Immunodeficiency Virus (HIV) and induces feline AIDS (Acquired Immunodeficiency Syndrome) in cats. Additionally, Chagas disease  and avian influenza have been modeled with SI models \cite{ChagasAgeInaba,SIavian}. 

Here we propose a graph dynamical model to study the evolution of disease that follow SI behavior. Simulating graphs have the disadvantage of being computationally intensive: an epidemic in a population of just a few hundred thousand nodes can take hours of computation to obtain averaged results over a range of model parameters. However, that is a once-off cost of doing away with the assumption of a fully-mixed population, and the range of possibilities it provides makes it a necessary trade-off. Mean-field approach may be used when large populations are simulated. 

There are several important studies that deal with epidemics using graphs \cite{GraphDrugAbuse,PartitionGraphControl,CentralTheoremEpidemicGraph}. On the other hand some interesting research using adaptative networks to model disease spread have been presented \cite{AnnealedGuerra,ReviewDiseaseAdaptative,contactchange}. For instance in \cite{SocialImpactNetwork} authors propose a interesting susceptible-infected-susceptible (SIS) epidemic model coupled with opinion dynamics to investigate the effects of social impact on the epidemic spreading in complex networks. The model consists of two types of nodes with different behavior patterns, active nodes and passive nodes.

The main aim in this article is to study the evolution of the SI epidemiological model and the creation of subgraphs due to the dynamic behavior of the individuals trying to avoid the contagious of the disease. The proposed dynamical graph model uses a single parameter which reflects the probability of rewire in order to avoid the disease. The numerical simulations are performed using Monte Carlo method in order to study the dynamics of the disease and the evolution of the subgraphs. This framework allows to analyze the simultaneous time evolution of the disease and the underlying graph topology. In this paper, we study the time evolution of the disease and the underlying graph topology assuming that the infectivity of the disease $\beta$ is coupled to the probability $p_{r}$ to relink to a different node (individual). We study the proposed graph model using different connectivity degrees, disease transmission rates and graph structure at the initial stage.

As usual we describe the stochastic contact graph with members of the population represented by vertices and with social contacts between individuals represented by edges. An edge can be seen as a social contact with the potential to transmit the disease if one of the nodes is infectious \cite{WittenNetwork}. The number of edges of a graph at a vertex is called the degree of the vertex. A well known definition is the degree distribution of a graph $p_{k}$, where $p_{k}$ is the fraction of vertices having degree $k.$ The degree distribution is fundamental in the description of the model using heterogeneous graphs. 

The layout of this paper is as follows. In Section $2$ the underlying classical compartmental mathematical model is introduced with a brief introduction to network modeling and the algorithms to simulate the disease evolution. Monte Carlo simulations results are performed in Section $3$. Discussion and conclusions are presented in Section $4$.

\section{The proposed SI dynamical graph model}

The classical $SI$ model considers a population with only two compartments, susceptible: S(t) and infected, I(t). The compartmental deterministic mathematical model can be represented analytically by the following nonlinear system of ordinary differential equations,
\begin{align}
\dot{S}(t)=& - \beta S(t) I(t), \notag \\
\dot{I}(t)=& \beta S(t) I(t), \label{1}  \\ 
\end{align}
where $\beta$ is the transmission rate due to the contact with infected individuals. The $SI$ model has many variants that include births, deaths, seasonal transmission rates or stochastic components \cite{Hethcote,Murray,ObesNetwork}. 

There are different approaches to simulate the evolution of a disease in a dynamical graph which depend on the assumptions to approximate the model to the real world. Here, we assume that the relink process is coupled to the intrinsic infectivity of the disease. It is important to remark that we do not included birth and death process since this is a first approach. Therefore, we assume a constant network size in all the Monte Carlo simulations. However, birth and death process may be included easily \cite{ObesNetwork}.

An important issue that need to be considered in the simulation of the networks, is that in the contagion process the value of the probability of
contagion per unit time $\beta$ needs to be modified for each $\Delta t$ in order to have an equivalent model to the continuous one \cite{WangNetworkEigen}. Here we are interested in studying the effect of an equivalent value of $\beta$ and we use a parameter $\hat{\beta}$ which it is not equal to the parameter $\beta$ of the continuous model (\ref{1}).

Here we present the steps of the algorithms that simulate the evolution of the epidemics in the dynamical graph. The number of nodes (individuals) and the total connections or links in the initial social network maintain invariant in the simulation. However, the connections between nodes changes over the simulation. It is clear that several different algorithms may be used for the process evolution. 

The main idea in this model is that the relink process is coupled with the disease transmission. In our proposed paradigm we assume that if the disease transmission is high an individual disregard the possibility of changing his/her relationships with infected individuals since she/he think that any way he/she will get infected. We study in this case the effect that has the relink process on the dynamics of the diseases. Additionally, we study the effect that has the connectivity degree $<k>$ and the initial network structure on the evolution of the epidemics in the coupled model.

The steps of the algorithm that simulates the evolution of the epidemics in our coupled dynamical graphs are the following:

\begin{itemize}
\item Create an initial graph with $N$ nodes where $(N-1)$ are susceptible and only one node is infected.
\item Advance one time step $\Delta t$ and choose one node randomly. 
\item If the selected node is susceptible then we select from his neighborhood nodes one randomly. If this node is infected then with probability $\hat{\beta}$ the susceptible node is infected and with probability $(1-\hat{\beta})$ the susceptible node break the contact and links with another node randomly. 
\item If the selected node is infected then we select from his neighborhood nodes one randomly. If this node is susceptible then with probability $\hat{\beta}$ the susceptible node is infected and with probability $(1-\hat{\beta})$ the susceptible node break the contact and links with another node randomly. If the selected node is infected the graph stays invariant.
\item Based on the previous steps the graph topology may change in each time step $\Delta t$. Then go to step $2$ and this process finish when there are not susceptible nodes in the graph or there are not active links, i.e. all the infected nodes are only linked to infected nodes or insulated.  
\end{itemize}

\section{Monte Carlo simulations of the $SI$ graph based model}

This section is devoted to present the simulation results for the coupled $SI$ graph model proposed. Simulations are made using the Monte Carlo method, assuming constant population and connections. Comparisons between different networks varying connectivity degree, disease transmission rate, relink probability and initial network structure are performed in order to investigate the effect that have these parameters on the transient and stationary states. Moreover the calculations are based upon different number of runs and population sizes. However, only some representative cases are shown. 

In order to investigate the dynamics of the SI graph it is necessary to reproduce the initial conditions of the population. Therefore, initially it is allocated only one infected individual in one of the communities in order to mimic the arrival of one infected individual to a region. In regard to the initial structure of the network we use three type of networks; random, regular and small world. It is important to notice that double connections and self-loops are not considered here. Additionally, the mean and standard deviations are computed.

\begin{figure}[htbp]
\centering
\begin{tabular}{cc}
\includegraphics[width=0.45\textwidth]{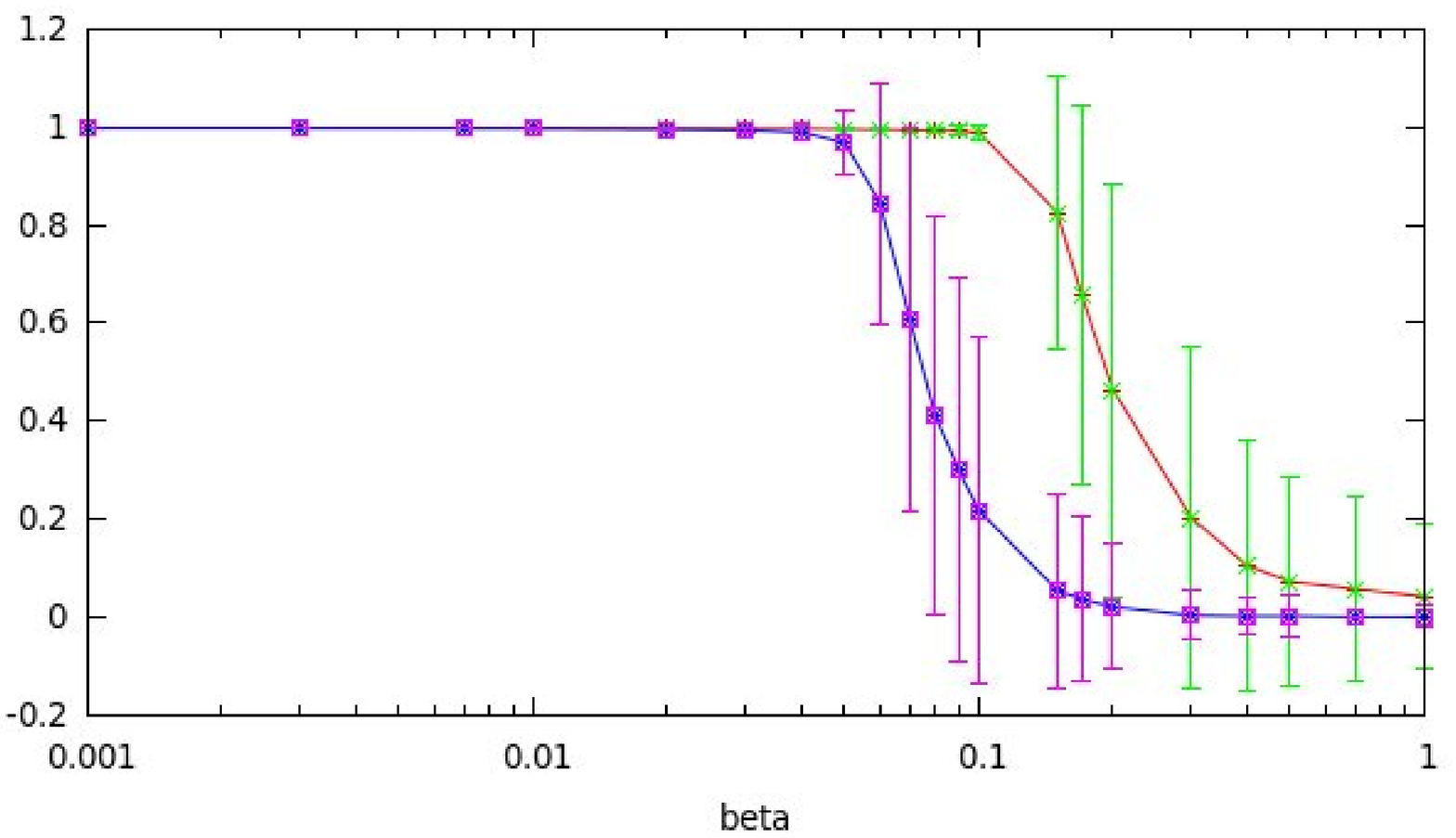}
& \includegraphics[width=0.45\textwidth]{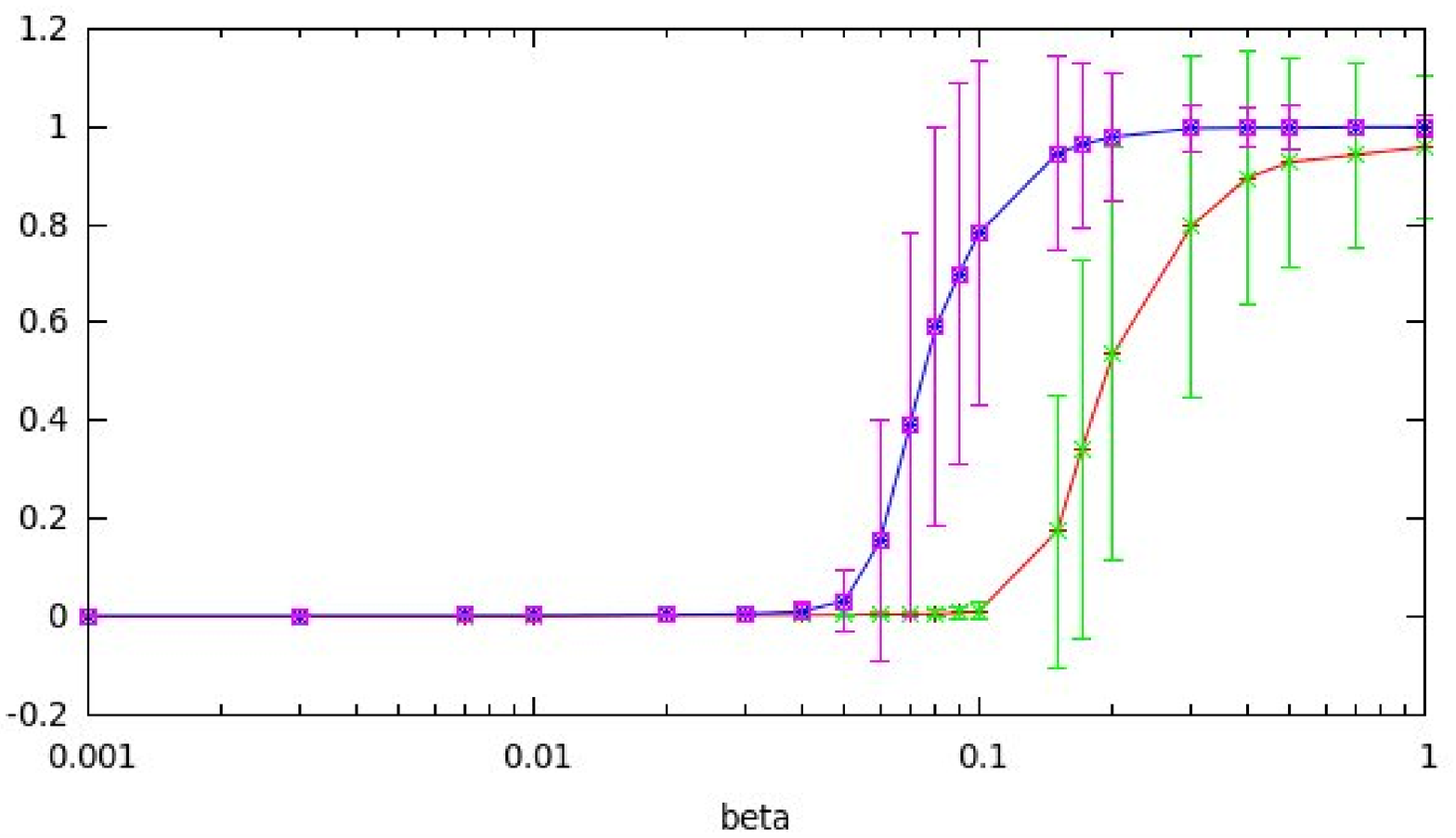} \\
\end{tabular}
\caption{Simulation of the epidemics varying $\hat{\beta}$ (transmission and relink). Proportion of susceptible individuals in the stationary state (Left side). Proportion of infected individuals in the stationary state (Right side).}
\label{fig8}
\end{figure}

The simulations are performed varying the probability of disease transmission $\hat{\beta}$. On the left hand side of Figure \ref{fig8} it can be seen that the proportion of susceptible individuals in the stationary state approaches to zero when the probability of disease transmission $\hat{\beta}$ increases (relink probability decreases). Additionally, there are less susceptible individuals at the stationary state for a connectivity degree $<k>=8$ since the disease reach more individuals. Analogously, on the right hand side of Figure \ref{fig8} it can be seen more infected individuals at the stationary state for a connectivity degree $<k>=8$. Notice that for approximately $\hat{\beta}<0.04$ the disease does not propagate since the infected node are isolated.

\begin{figure}[htbp]
\centering
\begin{tabular}{cc}
\includegraphics[width=0.45\textwidth]{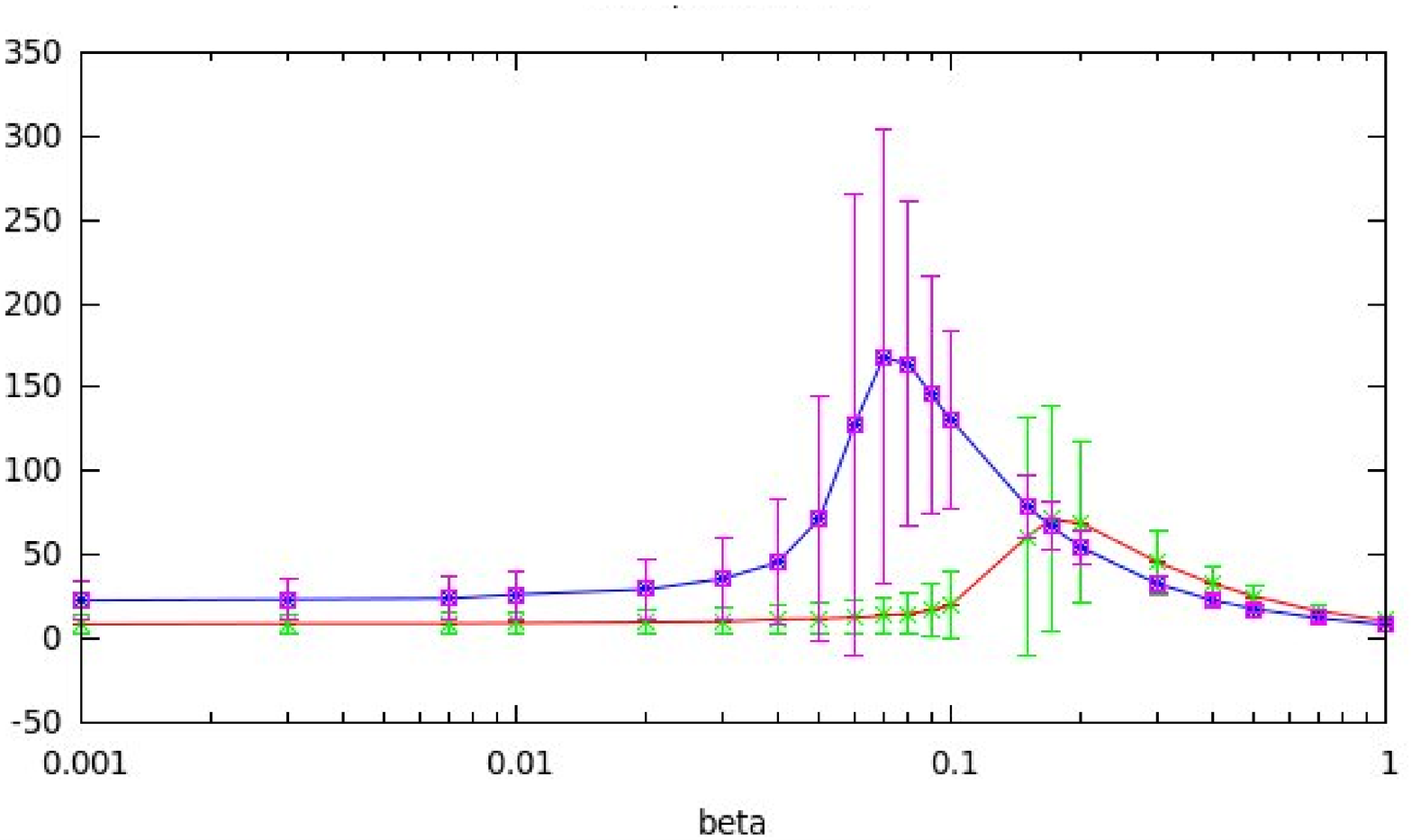}
& \includegraphics[width=0.45\textwidth]{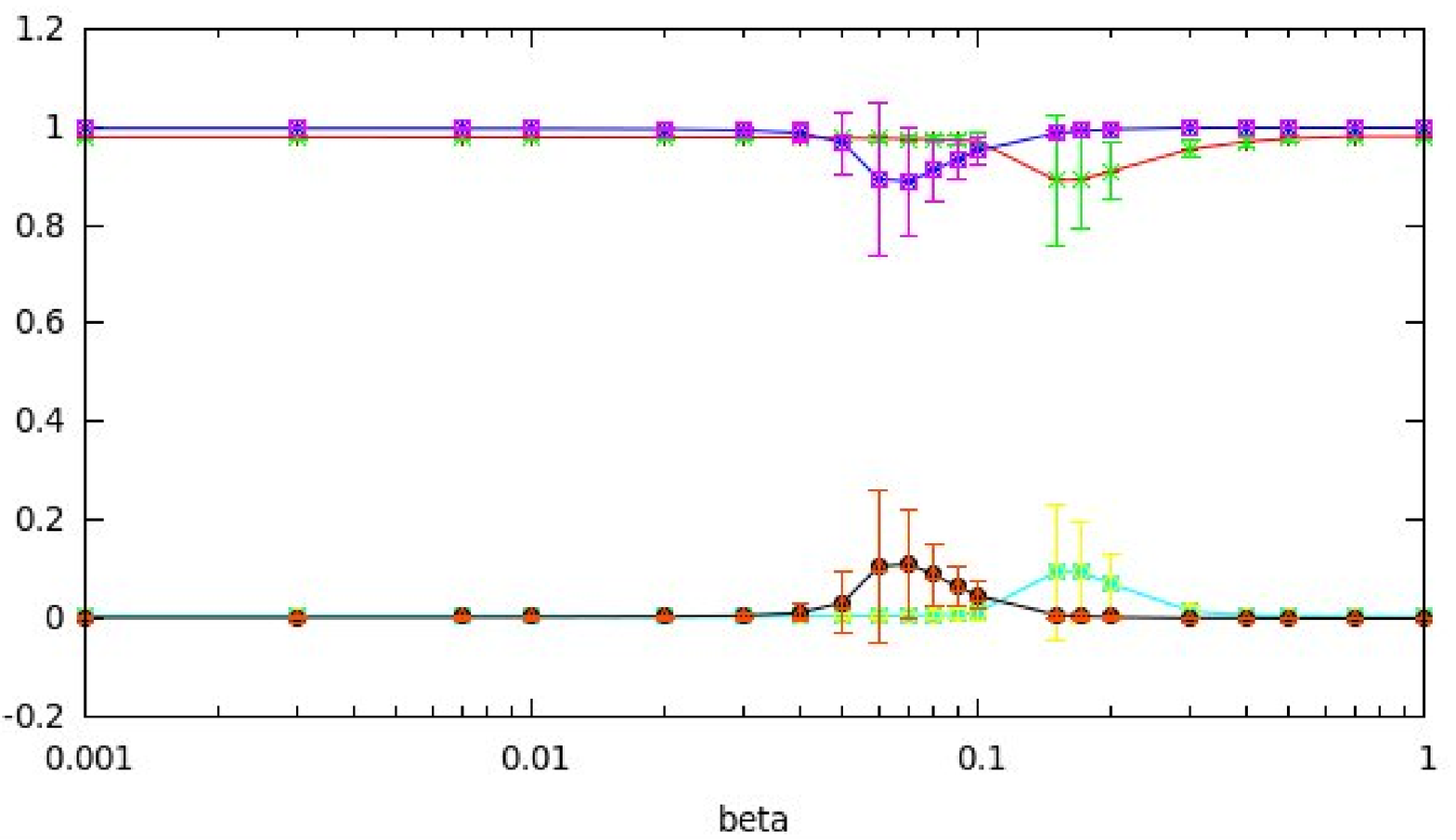} \\
\end{tabular}
\caption{Simulation of the epidemics varying $\hat{\beta}$ (transmission and relink). Time to reach stationary state (Left side). Size of the two biggest subgraphs when $<k>=4,8$. (Right side).}
\label{fig9}
\end{figure}

In regard to the extinction time of the epidemics it can be observed on the left hand side of Figure \ref{fig9} a very interesting behavior. At $\hat{\beta}\approx 0.06$ it is obtained a maximum extinction time for the network with connectivity degree $<k>=8$ and at $\hat{\beta}\approx 0.11$ for a connectivity degree $<k>=4$. This means that the disease spread easier on networks with higher connectivity. In addition, the maximum point may be interpreted as a point where some difficulties are presented for the disease to reach susceptible individuals in a fast way. For instance the infected nodes are almost isolated and cannot propagate the disease easy, but the dynamical graph still active. Thus, for larger values of $\hat{\beta}$ the disease reach more individuals in a faster way and for lower values of $\hat{\beta}$ the infected nodes are isolated faster. Thus, the main result obtained here is that when the probability of disease transmission $\hat{\beta}$ increases over a critical value $\hat{\beta}_{c}$ the extinction time decreases, which reflects that the graph is inactive (no more infectious).

One interesting result that helps to understand the dynamics of the epidemiological SI graph is presented on the right hand side of Figure \ref{fig9}. It can be seen that the second biggest subgraph in the network with connectivity degree $<k>=8$ has the biggest size at $\hat{\beta}=0.05$. On the other hand, the second biggest subgraph in the network with connectivity degree $<k>=4$ has the biggest size at $\hat{\beta}=0.109$. In the network with connectivity degree $<k>=8$ it has obtained a bigger second biggest subgraph. It is important to remark, that the points of maximum disjoint can be related to the points of maximum extinction time. This fact seems to have a logical explanation since a network with highest disjoint has more difficulties to propagate the epidemics in a easy way. Finally, it is important to mention that the region where is obtained the second biggest subgraph is close related to the highest extinction time. As we mentioned before this happens when the infected nodes are almost isolated and can not propagate the disease easy, but the dynamical graph still active. From a practical point of view, this fact is important for health institutions since they are interested in data regarding the time period of the epidemic. 

%

\section{Discussion and conclusions}

In this paper we studied the susceptible-infectious (SI) epidemiological model using dynamical graphs. Dynamical structures have been recently applied in many areas including complex systems. Dynamical graphs applied to epidemics consider generally that the nodes are individuals and the links represent different classes of relationships between individuals with the potential to transmit the disease. 

We studied the evolution of the SI epidemiological model and the creation of subgraphs due to the dynamic behavior of the individuals trying to avoid the contagious of the disease. In the proposed dynamical graph model we use a single parameter which reflects the probability of rewire that represent actions to avoid the disease. In this way we assumed that the infectivity of the disease $\beta$ is coupled to the probability $p_{r}$ to relink to a different node (individual). We studied the proposed graph model using different connectivity degrees, disease transmission rates and graph structure at the initial stage.

The numerical simulations using Monte Carlo method show that the dynamical behavior of individuals affects the evolution of the subgraphs in the coupled SI graph model. Numerical results show that the proportion of susceptible individuals in the stationary state approaches to zero when the probability of disease transmission $\hat{\beta}$ increases (relink probability decreases) as was expected. However, the main result obtained is that when the probability of disease transmission $\hat{\beta}$ increases over a critical value $\hat{\beta}_{c}$ the extinction time decreases, which reflects that the graph is inactive (no more infectious).

In addition, we obtained that the region of maximum disjoint of the graph can be related to region of maximum extinction time. This fact seems to have a logical explanation since a network with highest disjoint has more difficulties to propagate the epidemics in a easy way. This fact can is due to that the infected nodes are almost isolated and cannot propagate the disease easy, but the dynamical graph still active. From a practical point of view, this fact is important for health institutions since they are interested in data regarding the time period of the epidemic.

We can conclude that the rewire probability changing the structure of the network has an important influence on the evolution of the epidemics. As expected the rewire probability protects more individuals from the society against the disease, but not necessarily a higher rewire probability means that the epidemic extinct faster. Moreover, the results obtained in this paper show that the transmission disease and network topology, play an important role in determining the evolution and outcome of a particular epidemic scenario on an adaptive network.

%
\bibliographystyle{unsrt}
\bibliography{BibNetworkDynamical}

\end{document}